%% file: ms.tex
\documentclass[manuscript]{aastex}
\usepackage{epsfig}
\usepackage{rotating}

\slugcomment{To appear in AJ}

\shorttitle{Photometry of SIM Quasars}
	
\shortauthors{Ojha et al.}

\begin{document}

\title{Photometric observations of selected, optically bright quasars 
    for Space Interferometry Mission and other future celestial reference frames}

\author{Roopesh Ojha\altaffilmark{1},\altaffilmark{6}}
\affil{NVI/United States Naval Observatory, Washington, DC 20392 }

\author{Norbert Zacharias\altaffilmark{2},\altaffilmark{6}}

\author{Gregory S. Hennessy\altaffilmark{3},\altaffilmark{6}}

\author{Ralph A. Gaume\altaffilmark{4}}

\and 

\author{Kenneth J. Johnston\altaffilmark{5}}
\affil{United States Naval Observatory, Washington, DC 20392 }

\altaffiltext{1}{rojha@usno.navy.mil}
\altaffiltext{2}{nz@usno.navy.mil}
\altaffiltext{3}{gsh@usno.navy.mil}
\altaffiltext{4}{rgaume@usno.navy.mil}
\altaffiltext{5}{kjj@usno.navy.mil}
\altaffiltext{6}{Visiting astronomer, Cerro Tololo Inter-American 
Observatory, National Optical Astronomy Observatory, which are operated 
by the Association of Universities for Research in Astronomy (AURA), 
under cooperative agreement with the National Science Foundation (NSF).}

\begin{abstract}
Photometric observations of 235 extragalactic objects that are
potential targets for the Space Interferometry Mission (SIM) are
presented. Mean {\it B}, {\it V}, {\it R}, {\it I} magnitudes at the 5\% level are
obtained at 1 - 4 epochs between 2005 and 2007 using the 1-m
telescopes at Cerro Tololo Inter-American Observatory and Naval
Observatory Flagstaff Station. Of the 134 sources which have V
magnitudes in the Veron \& Veron-Cetty catalog a difference of over
1.0 mag is found for the observed$-$catalog magnitudes for about 36\%
of the common sources, and 10 sources show over 3 mag
difference.  Our first set of observations presented here form the
basis of a long-term photometric variability study of the selected
reference frame sources to assist in mission target selection and to
support in general QSO multi-color photometric variability studies.
\end{abstract}


\section{Introduction}\label{intro}
The Space Interferometry Mission (SIM) is a proposed facility of the National 
Aeronautics and Space Administration (NASA) that will be the first space-based
Michelson interferometer for astrometry.  
The version of SIM described here is SIM-Lite \citep{sim-lite}.
It will have a 6 m baseline and operate in the optical/near-IR waveband. 
For stars brighter than $V = 20$ it will deliver a global 
astrometric accuracy of 4 $\mu$as. When operating in its narrow angle mode, it 
will achieve a positional accuracy of 1.0 $\mu$as for a single measurement,
with even smaller errors for differential positional accuracy at the end of 
five years (the nominal mission lifetime). This performance is at least 2 orders of 
magnitude better than any existing instrument. Astrometry 
at these unprecedented levels of precision will significantly impact a broad 
range of astronomy, from the search for planetary systems to a more accurate 
value of the Hubble constant, from measurements of dynamical masses of binary 
stars to the dynamics of accretion disks around supermassive black holes 
\citep{unwin08}.

A key contribution of SIM will be the creation of a new absolute reference frame.
The International Celestial Reference Frame (ICRF) is currently the fundamental 
celestial reference frame and the standard frame for all astrometry. It is defined 
by the radio positions of 212 extragalactic radio sources with most having errors 
below $1$ mas \citep{johnston95, ma98}. SIM will be capable of defining a celestial 
reference frame at optical/near IR wavelengths, which will have orders of magnitude 
greater accuracy than the ICRF with an expected level of about $4 \mu$as.
The SIM reference frame will be defined by the positions and motions of 1304 
reference grid stars uniformly distributed over the entire sky. 
The SIM grid stars are selected K0III stars of visual magnitude 10-12.
The positions of these grid stars will be determined (end of mission) relative 
to one another to an accuracy of $< 4 \mu$as in position and parallax and 
$< 4 \mu$as $\mathrm{yr}^{-1}$ in proper motion, with individual observations 
being accurate to $10\mu$as. 

In order to remove any residual rotation in the SIM stellar reference frame 
i.e.~to make the frame quasi-inertial, and enable the determination of absolute 
proper motions, a number of extragalactic sources will be observed by SIM.  
These distant sources are assumed to have negligible proper motions and 
parallax. In theory, a minimum of two fixed extragalactic objects need to be 
observed for this purpose. In practice, possible structure and variability of the 
extragalactic sources and zonal errors in the SIM stellar grid solution dictate 
that a global distribution of at least 20 - 50 extragalactic sources be used
\citep{valeri_qso}. 
Besides making the SIM reference frame inertial for dynamical applications,
the orientation of this new reference frame will be aligned to the current
ICRF by observing a subset of
the ICRF extragalactic sources which display bright, optical counterparts.

SIM is a pointed mission and only a limited number of targets can be observed. 
Although targets as faint as 20th magnitude can be observed with SIM to 
full accuracy the required observing time is prohibitive for a large number
of targets.  We wish to minimize observing time for SIM by finding the 
brightest, suitable sources possible and by ensuring a good all-sky distribution 
required to establish an inertial reference frame for SIM global astrometry.
The goal is to select mainly {\it R} $\le$ 16$^{m}$ QSO targets with some fainter
sources to fill in gaps in the sky coverage.

Here, we present the results from photometric observations of potential
SIM quasar targets in the Johnson {\it B}, {\it V}, {\it R}, and {\it I} bands. This 
program is a
long-term effort and we plan on continued observations to study variability.
Color information is required to be able to derive the expected brightness
of a target in the not yet specified SIM instrumental bandpass.

Of course these observations will also serve the wider astronomical
community.  For example, the fully funded J-MAPS mission \citep{gaume09} 
whose goal is to generate a nearly 40 million star catalog with better than $1$ 
mas positional accuracy and photometry to the $1 \%$ accuracy level 
or better, will also need to observe quasars. Observations of over $100$ quasars 
(which being at large distances have quasi-zero parallax) will be needed in order 
to make the J-MAPS parallaxes absolute and to minimize zonal parallax errors. Due 
to their quasi-zero proper motions observations of quasars will also be needed to 
make the J-MAPS coordinate system inertial so that J-MAPS astrometry will be relevent 
to any dynamics study. Finally, observations of quasars will be needed to align 
the J-MAPS coordinates to the standard system (ICRF) at the mean epoch of 
observations. Such an alignment is essential for direct positional comparisons 
of targets common to the J-MAPS optical and the ICRF radio systems. 

Finally, despite lots of recent new observations \citep{wilhite08, bachev09, bauer09}, 
quasar variability remains a poorly understood phenomenon 
and our dataset will allow us to begin addressing questions such as 
the mechanisms that give rise to quasar variability (e.g., see 
\citet{vandenberk04} and references therein).

\section{Observations}\label{observe}
With the ultimate goal of selecting suitable, compact, optically bright QSOs 
for a future astrometric celestial reference frame which will be established 
by a mission like SIM, 242 extragalactic sources preferrably brighter than 
$16^{th}$ visual magnitude and with no noticeable asymmetric structure on 
the Digitized Sky Survey images as compared to nearby stellar objects were 
chosen by visual inspection from the Veron-Cetty catalog (\citet{veron06}; 
hereafter VCV06). Photometric observations of these sources were carried out 
in the Johnson {\it B}, {\it V}, {\it R} and {\it I} bands and results for 214 of these 
sources are presented along with an additional 21 sources which were observed 
as part of related observing programs. 

Objects in the northern hemisphere were primarily observed with the Naval 
Observatory Flagstaff Station (NOFS) 1.0 m Ritchey-Chreti{\`e}n reflector,
using a 2k CCD camera during four observing runs between 2005 and 2007.
Objects in the southern hemisphere were observed with the Cerro Tololo 
Inter-American Observatory (CTIO) Small and Moderate Aperture Research 
Telescope System (SMARTS) 1.0 m in Chile. The 4kx4k pixel camera 
Y4KCam, was used and four successful observing runs were carried out from 
2005 through 2007. 
The epochs and lengths of all successful observing runs with both 
telescopes are summarized in Table~\ref{observesummary}. 
Standard photometric calibration observations were performed as part of
these runs, including dome flats and observations of Landolt calibration 
stars \citep{landolt92}.

\section{Data Reduction}\label{reduce}

\subsection{Raw Data Processing}

For the CTIO 1-m data, modified scripts based on the 4k CCD processing 
pipeline developed by P. Massey 
\footnote{http://www.lowell.edu/users/massey/obins/y4kcamred.html}
in IRAF \footnote{IRAF is 
distributed by the National Optical Astronomy Observatory, which is operated 
by the Association of Universities for Research in Astronomy, Inc.,  under 
cooperative agreement with the National Science Foundation.} 
were used for overscan, trim, bias and flat-field operations (four-output system 
from Yale University). Standard IRAF routines were used for the raw data 
processing from the single output 2k CCD used at the NOFS 1-m telescope. 
All individual bias and flat frames were looked at and only acceptable data 
were used for combining.

\subsection{Astrometric Data Processing}

All processed object frames were examined visually and the QSO target 
was identified with respect to finder charts. Basic statistical information 
and remarks were entered in a quality control table. The astrometric pipeline 
of the radio-optical reference frame link program \citet{zacharias05} 
was adopted for processing of both the CTIO 1-m and NOFS 1-m data. 
Image centroids were obtained from two-dimensional Gaussian profile least-squares 
fits to the pixel counts of detected sources. The UCAC2 catalog 
\citep{zacharias04} provided the reference stars for the astrometric reductions. 
An eight-parameter plate model was used for astrometric reduction after determining 
and applying corrections for atmospheric refraction and mean optical field angle 
distortion.

There are dedicated, high accuracy, astrometric observations going on at the 61inch 
telescope at NOFS \citep{zacharias08a, zacharias08b} with results to be published in
2010 for about 12 sources on the few milliarcsecond level.  A set of over 200 QSO's 
has been observed at somewhat lower accuracy with respect to UCAC3 reference 
stars, which will be published in 2009. 

\subsection{Photometric Data Processing}

Instrumental magnitudes were obtained from aperture photometry performed on 
each detected object as part of the astrometric pipeline. Pixel counts in the 
background ring were sorted and outliers excluded before deriving the mean 
local background. Photometric standard fields were observed typically 3 - 5 times 
a night at various air masses.  Photometric calibration per filter 
was performed with the Landoldt stars \citep{landolt92} and a two or three-parameter
model determining the zero point and linear terms as a function of airmass 
and color (if needed), respectively, was constructed.
The decision to include a color term was based on results from
test reductions for each filter and night, at which time also outliers
and potential problems were identified.  Least-square fit 
results and residuals were examined to determine the photometric quality of 
the night. The $1 \sigma$ error on the photometric fit had a median value of 35, 32, 
37 and 35 mmag, and the extinction had a median value of $-0.268$, $-0.155$, 
$-0.105$ and $-0.056$ for the {\it B}, {\it V}, {\it R}, and {\it I} bands, respectively. Color 
coefficients were 
not used for $60\%$ of our nights. Their typical value, when used, was $-0.058$. For 
acceptable data the derived photometric calibration parameters were applied to the 
QSO target observations to obtain individual {\it B}, {\it V}, {\it R}, and {\it I} magnitudes.

Color information of the QSO targets were initially not available and a mean 
color was adopted for the first photometric calibration step.  After deriving 
preliminary magnitudes in all bands, colors were derived and iteratively used 
in the photometric calibrations for subsequent processing to arrive at the 
final target magnitudes.

\section{Results}\label{results}

\subsection{Distribution of Targets}

\begin{figure}
\epsscale{.60}
\plotone{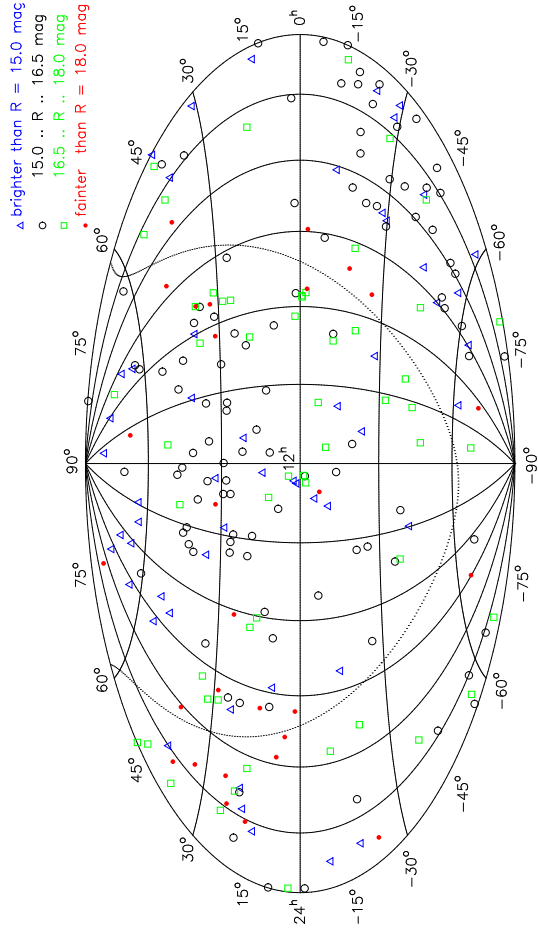}
\caption{Sky distribution (Aitoff plot) of observed sources. 
  Symbol types indicate {\it R}-magnitude. \label{aitoff}}
\end{figure}

Figure~\ref{aitoff} shows the observed sources in the sky, coded by 
symbol for groups of {\it R}-magnitude. The plot shows that the sources are 
well distributed over the sky; however, the area near the galactic plane 
(as expected) is significantly more sparsely populated than the rest of the sky. 
In sky areas with a large number of sources, a reference frame target 
source can be selected from a number of candidates depending on further 
investigations.  For areas close to the galactic plane the choices are 
limited and often only a target of 18th or fainter magnitude is 
available. However, even if a relatively bright target cut-off limit has 
to be adopted, the sky distribution of sources is sufficiently homogeneous
to support a highly accurate future reference frame
(Makarov, private communication).
 
\subsection{Mean Magnitudes}

Individual magnitudes of our target sources were collected and averaged 
(per filter) for multiple observations over the period of an observing run, 
which typically extended to a few nights. In a few cases a larger than 
expected scatter was observed. From the combined data, a fraction of 
the nights were identified with non photometric conditions and all data points 
obtained during that period of time were excluded. 
The final results for each object and each observing run are 
presented in Table~\ref{finalresults}. In this table, the first 
column has the J2000.0 name of the target, columns 2 - 5 have 
the mean magnitudes per run for {\it B}, {\it V}, {\it R}, and {\it I} filters, 
respectively. Columns 6 - 9 have the respective standard errors per epoch, a 
value of  '\ldots'  indicates no measurement through that particular filter. 
 Columns 10 - 13 show the ({\it B} - {\it V}),  ({\it V} - {\it R}), ({\it R} - {\it I}) and 
 ({\it V} - {\it I}) colors respectively. Columns 14 - 17
show the number of times the object was observed with a particular filter.
The final column records the epoch label (observing run, see Table~1).

\subsection{Colors}

\begin{figure}
\epsscale{1.00}
\plotone{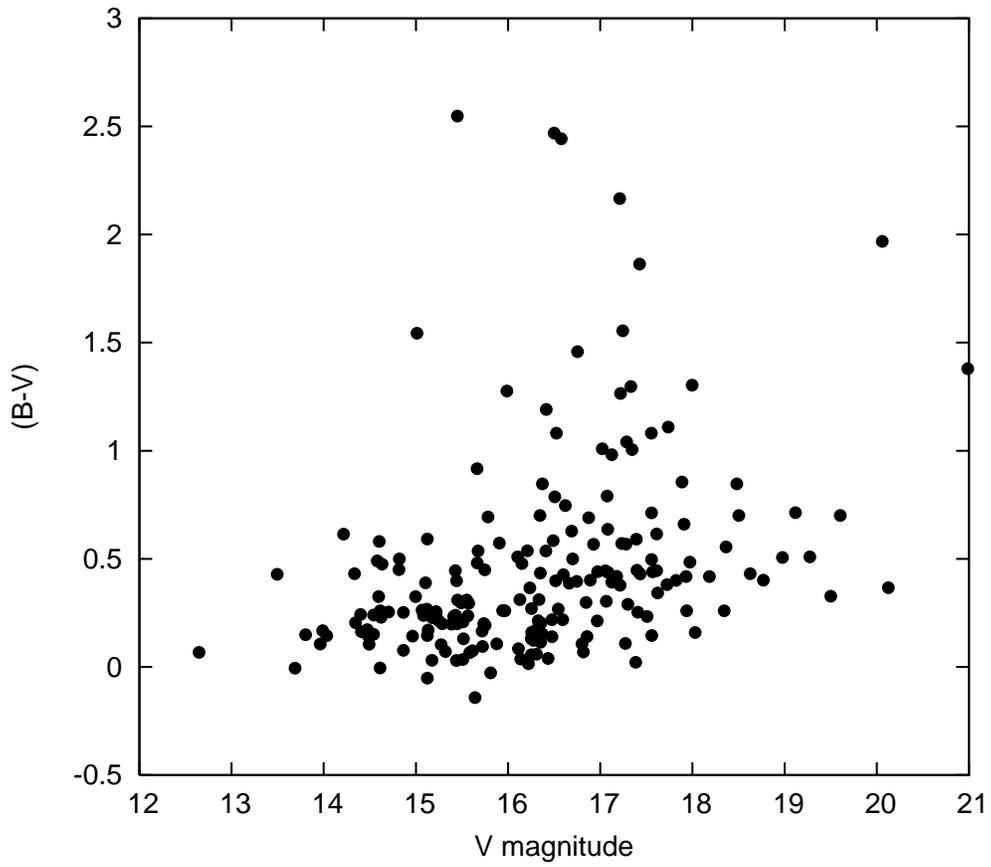}
\caption{Plot of observed ({\it B} - {\it V}) color vs. observed {\it V}-magnitude. 
  \label{BVvsV}}
\end{figure}

\begin{figure}
\plotone{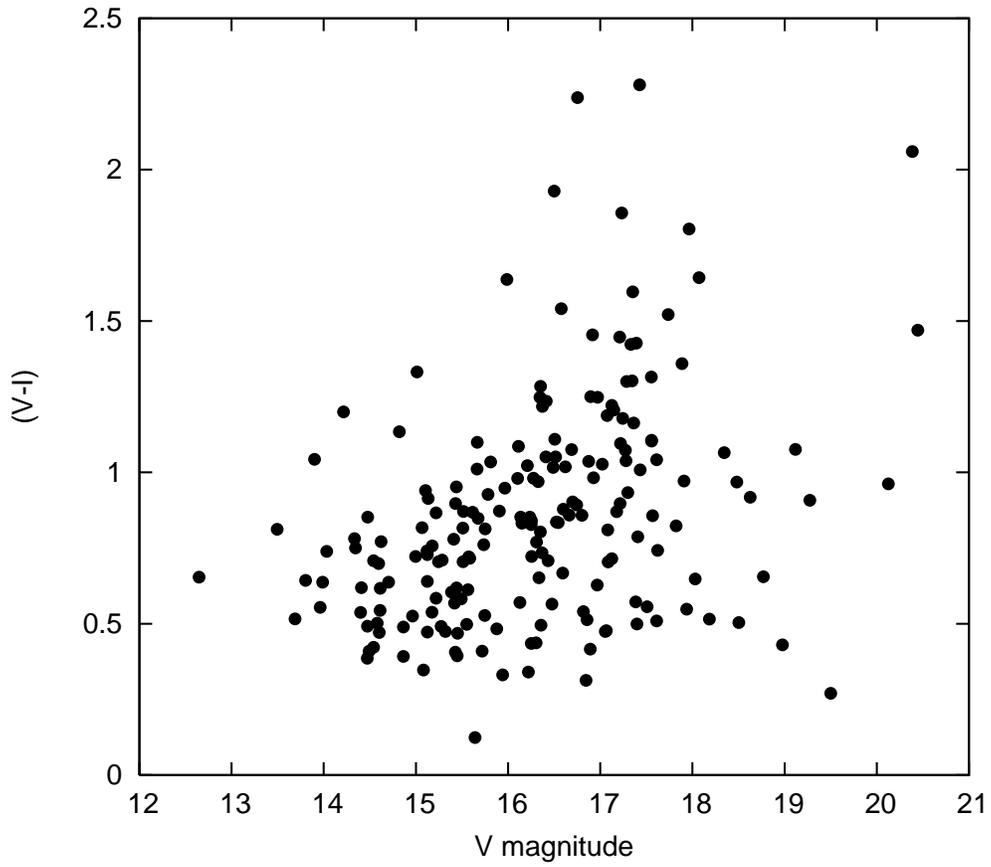}
\caption{Plot of observed ({\it V} - {\it I}) color vs. observed {\it V}-magnitude.
  \label{VIvsV}}
\end{figure}

\begin{figure}
\plotone{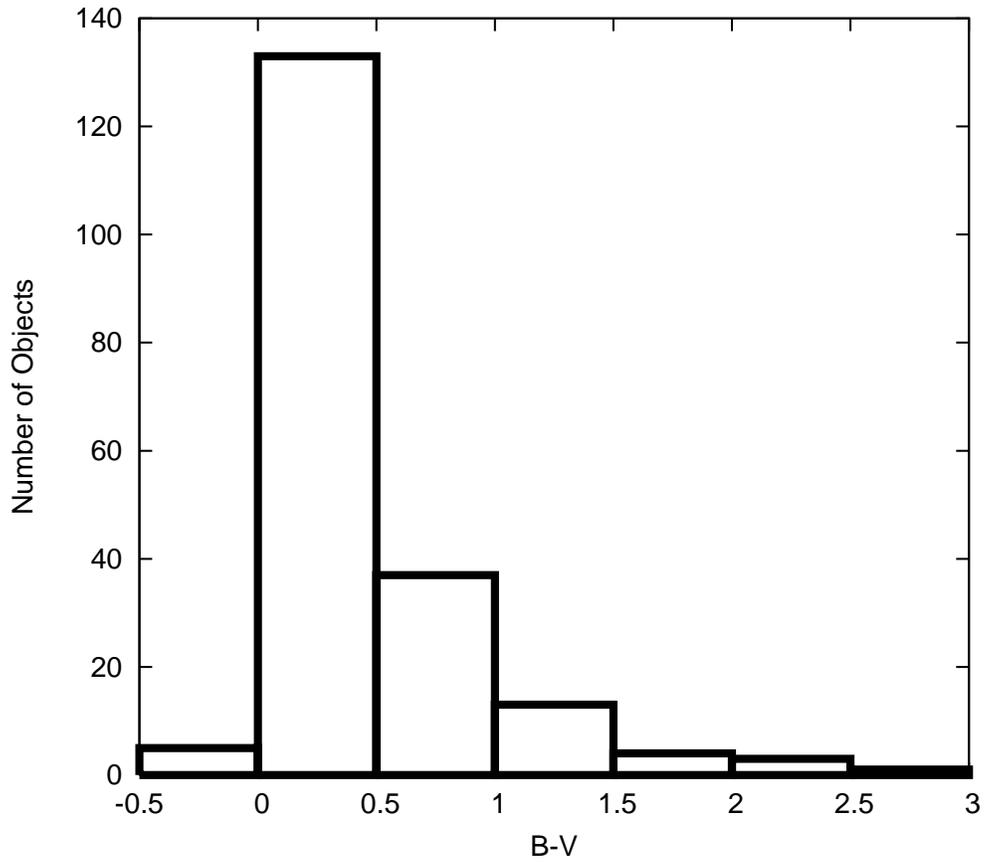}
\caption{Histogram of observed ({\it B} - {\it V}) color distribution. \label{B-Vhisto}}
\end{figure}

Figure~\ref{BVvsV} and Figure~\ref{VIvsV} show the observed ({\it B} $-$ {\it V}) 
and ({\it V} $-$ {\it I}) colors, respectively. A large range of colors is observed 
with five targets having ({\it B} $-$ {\it V}) $\le$ 0, 170 targets in the 0 - 1 range 
and 21 targets in the $\ge$ 1 range, respectively (Fig~\ref{B-Vhisto}). This is 
expected for a sample of QSOs with a large spread in redshift. 
This color knowledge is important for future reference frame target 
selection.  To optimize the expensive integration time of missions like 
SIM a prediction of the target brightness in the instrumental system will 
be required on the few tenths magnitude level.  Our {\it B}, {\it V}, {\it R}, and 
{\it I} results 
span the expected bandpass of the SIM mission to allow this brightness 
estimate, however, more observations near the mission epoch will be 
required to take variability into account.

\subsection{Comparison with Veron}

\begin{figure}
\plotone{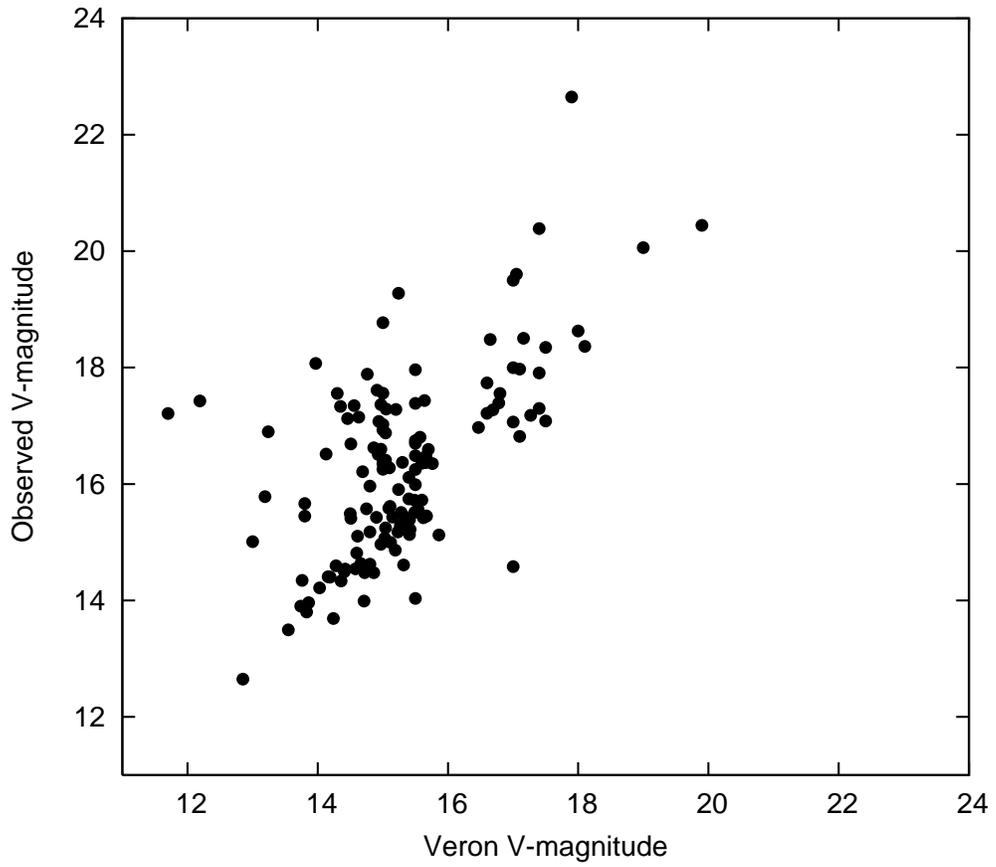}
\caption{Plot of observed {\it V}-magnitudes vs. Veron catalog {\it V}-magnitudes. 
  \label{veroncompare}}
\end{figure}

\begin{figure}
\plotone{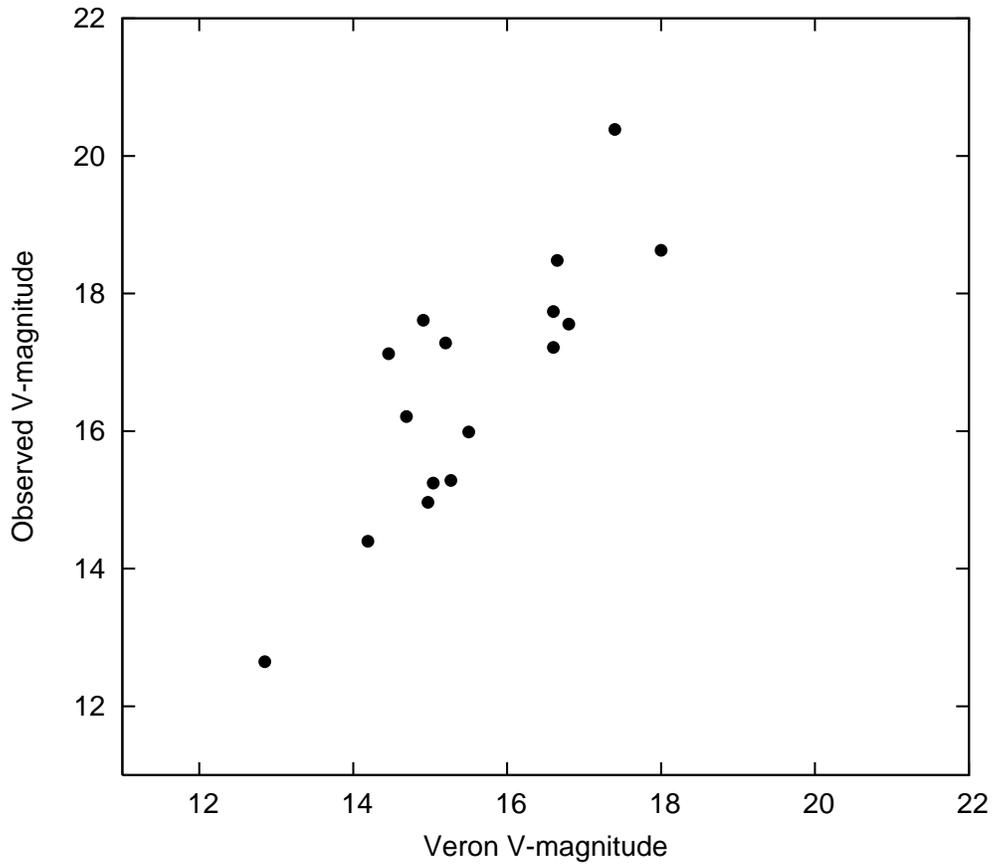}
\caption{Plot of observed {\it V}-magnitudes vs. Veron catalog {\it V}-magnitude 
  for those sources observed at three or more epochs. 
  \label{veroncomparemultiple}}
\end{figure}

\begin{figure}
\plotone{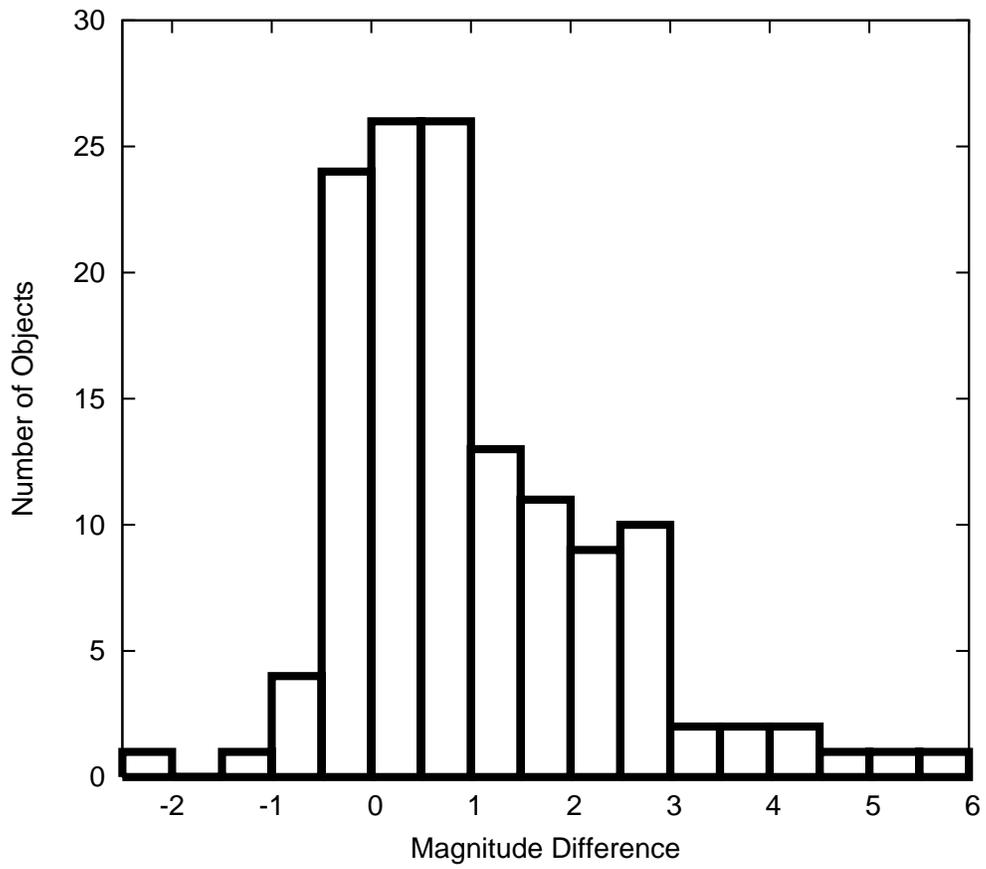}
\caption{Histogram of magnitude differences between observed {\it V}-magnitudes 
  and Veron catalog {\it V}-magnitudes. A positive difference indicates a larger 
  observed {\it V}-magnitude (fainter) as compared to the Veron value.  
  \label{magdiffshisto}}
\end{figure}

Figure~\ref{veroncompare} shows our observed {\it V}-magnitudes compared 
with {\it V}-magnitudes from the VCV06 for the 134 sources where they were available. 
Figure~\ref{veroncomparemultiple} shows our observed {\it V}-magnitudes 
for those 16 sources which were observed at three or more epochs and 
for which VCV06 V-magnitudes were available.
Both Figure~\ref{veroncompare} and Figure~\ref{veroncomparemultiple} 
show large differences between the Veron-Cetty and our observed magnitudes, 
sometimes exceeding 2 mag in either direction. 
The histogram in Figure~\ref{magdiffshisto} shows the distribution 
of V-magnitude difference between our observations and the Veron catalog. 
The internal errors of our observations are typically below 0.1mag
while the errors of the Veron magnitudes are typically not known.
Still differences of more than 2 mag suggest    
 mainly not a random observational error issue, rather being 
a physical change in brightness of the targets. This underlines the 
requirement for more, dedicated observations at an epoch of interest 
and clearly shows that a brightness look-up in some published table is not 
adequate for the task of scheduling SIM mission time to observe those targets
or even to select targets to a desired limiting magnitude.

\subsection{Comparison with SDSS}

We also made a comparison of our results with a dedicated sample, the Sloan Digital 
Sky Survey (SDSS DR5; \citet{adelman07}). Using a list of our object right ascension and 
declinations in decimal degrees we made an ``object cross id" type of search of the SDSS 
DR5 which yielded $\sim 85$ matches. We then checked that the positions of these matches 
were within $1$ arcsec of our quasar positions and were left with a list of $71$ matches 
we could use for comparison. 

\begin{figure}
\plotone{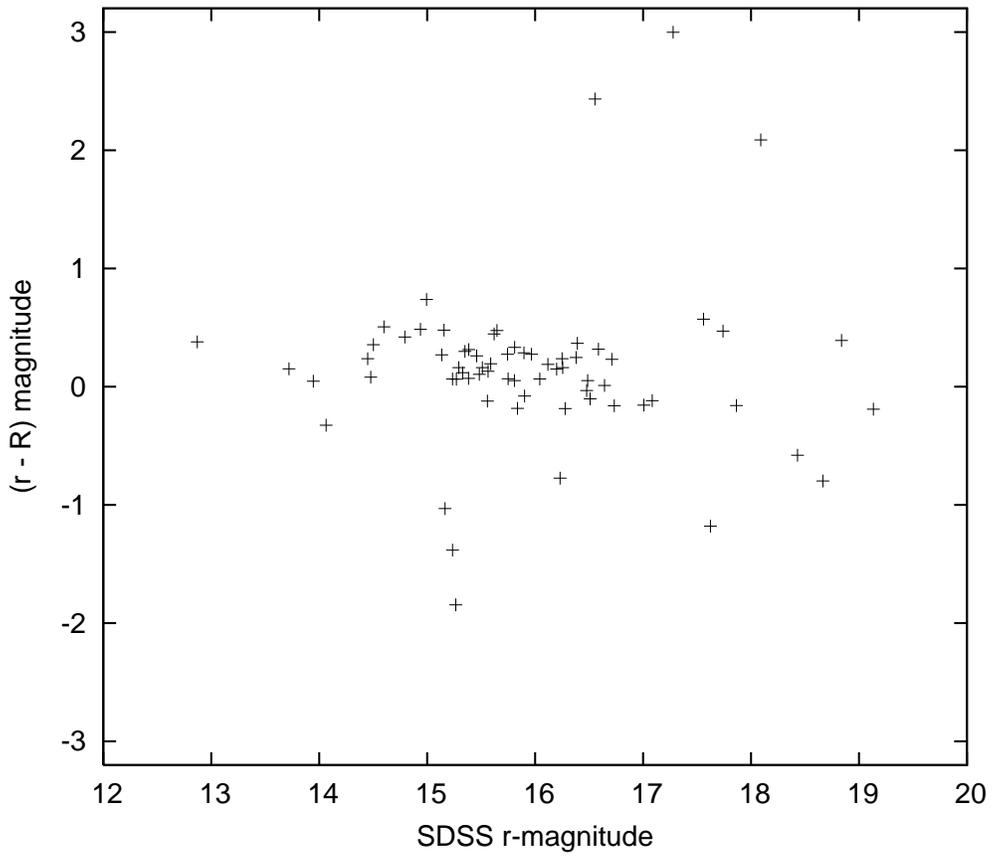}
\caption{Plot of ({\it r} - {\it R}), the difference of the SDSS {\it r} -magnitude and observed 
{\it R}-magnitude, against SDSS {\it r}-magnitude  
  \label{SDSS_r-Rvsr}}
\end{figure}

\begin{figure}
\plotone{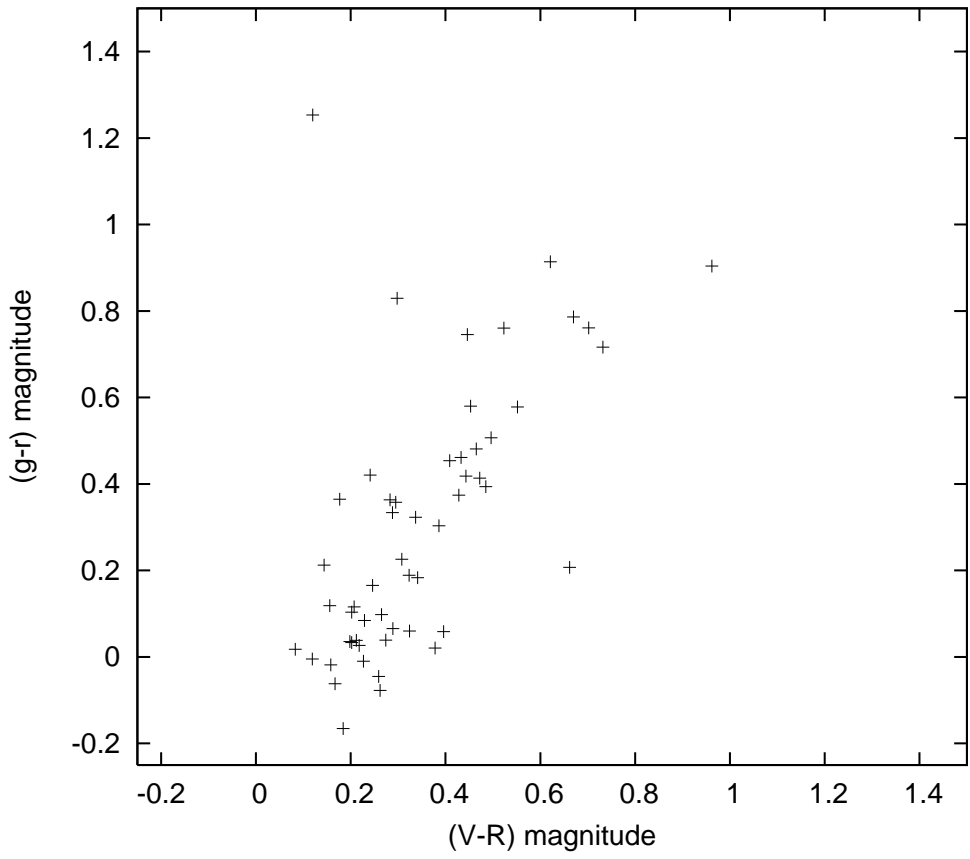}
\caption{Color-color plot of SDSS ({\it g} - {\it r}) against observed ({\it V} - {\it R}). 
  \label{SDSS_color-color}}
\end{figure}

Figure~\ref{SDSS_r-Rvsr} is a plot of ({\it r} - {\it R}), the difference of the SDSS {\it r} -magnitude 
and our {\it R}-magnitude, against SDSS {\it r} -magnitude for the 71 sources they were 
available. It shows relatively small differences between our observed {\it R}-magnitudes and 
the SDSS {\it r} -magnitudes down to {\it r} -magnitudes of about 19. The three sources fainter 
than this were outliers and are not included in the plot. It is possible that these three are 
mismatches. Figure~\ref{SDSS_color-color} shows a color-color plot of the SDSS ({\it g} - {\it r}) 
against our ({\it V} - {\it R}). Two outliers, both with very large negative ({\it g} - {\it r}) value, are 
not shown. This plot shows 
good agreement between the SDSS magnitudes and those presented here.

\subsection{Variability}

All our 235 sources are observed at least once in most of the filters.  
Many sources have multiple epoch observations. There were 24 sources 
that were observed in 3 or more observing runs and the scatter 
in their mean magnitudes were calculated and are shown in 
Table~\ref{magvariability} together with their overall mean magnitudes 
i.e.~averaged over all observing runs, and the largest formal error of 
a magnitude per observing run.
The formal error and the scatter columns are directly comparable
to look for a possible, physical variability of a source.
However, at this point we have only few epochs of a light curve
and we are dealing with small number statistics.

\section{Discussion}\label{discuss}

The goal of about 5\% photometric error per epoch was achieved for most 
sources, which is sufficient for the purpose of predicting the brightness 
of our targets in any instrumental system between {\it B} and {\it I} on the 10\% 
level to optimize future mission target selection and integration time.

Internal, photometric errors were calculated for single observations.
The errors given are the RSS (root-square-sum) of an assumed systematic
error floor of 0.01 mag, the Poisson noise error, and the nightly fit 
error to the standard stars.  Most targets are well exposed with low
random noise errors (about 0.01 mag) thus the nightly fit error
dominates the photometric error of individual observations.

Table~\ref{formalerrors} lists results for cases with formal error 
estimates from more than 1 observation per epoch and filter. Columns 
2 and 3 show the total number of such cases (per filter) and the number 
of cases with a formal photometry error (1$\sigma$) of less than or equal
to 5\%.  The last two columns give the median and largest formal error of
each sample, respectively. This shows that our goal of reaching 
a few percent precision has been achieved on most targets.

\begin{figure}
\plotone{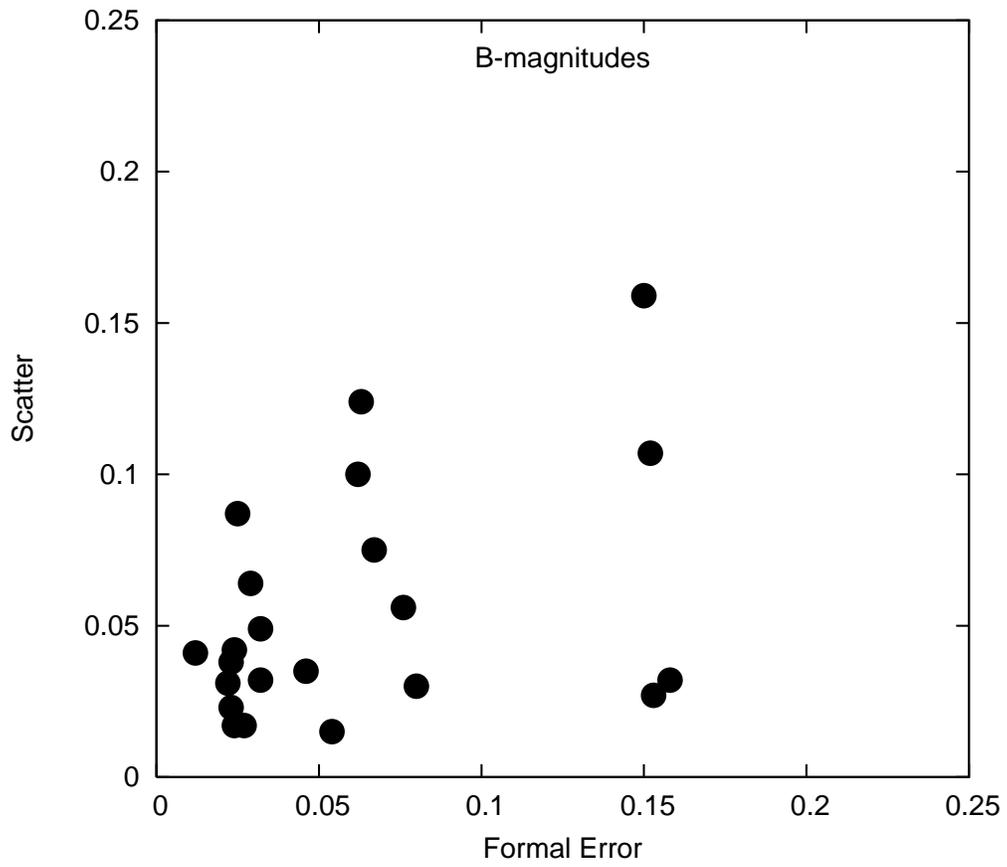}
\caption{Scatter in the {\it B} band data as a function of largest formal error.  
  \label{Berrorvsscatter}}
\end{figure}

\begin{figure}
\plotone{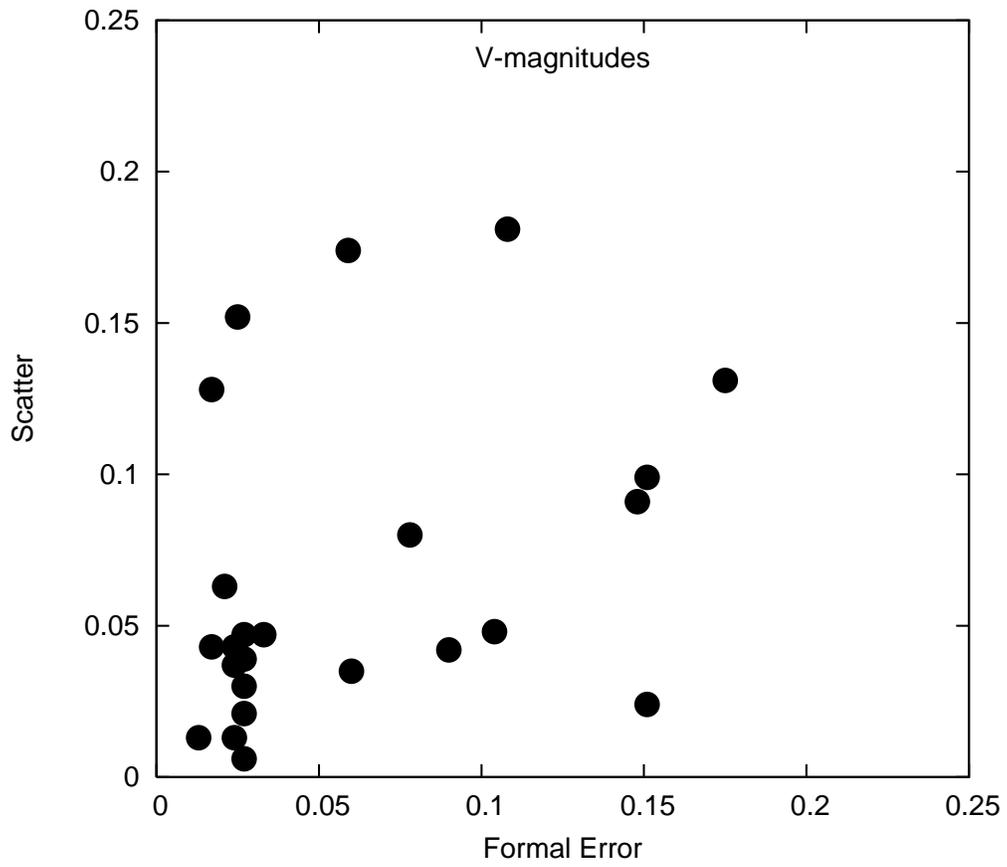}
\caption{Scatter in the {\it V} band data as a function of largest formal error.  
  \label{Verrorvsscatter}}
\end{figure}

\begin{figure}
\plotone{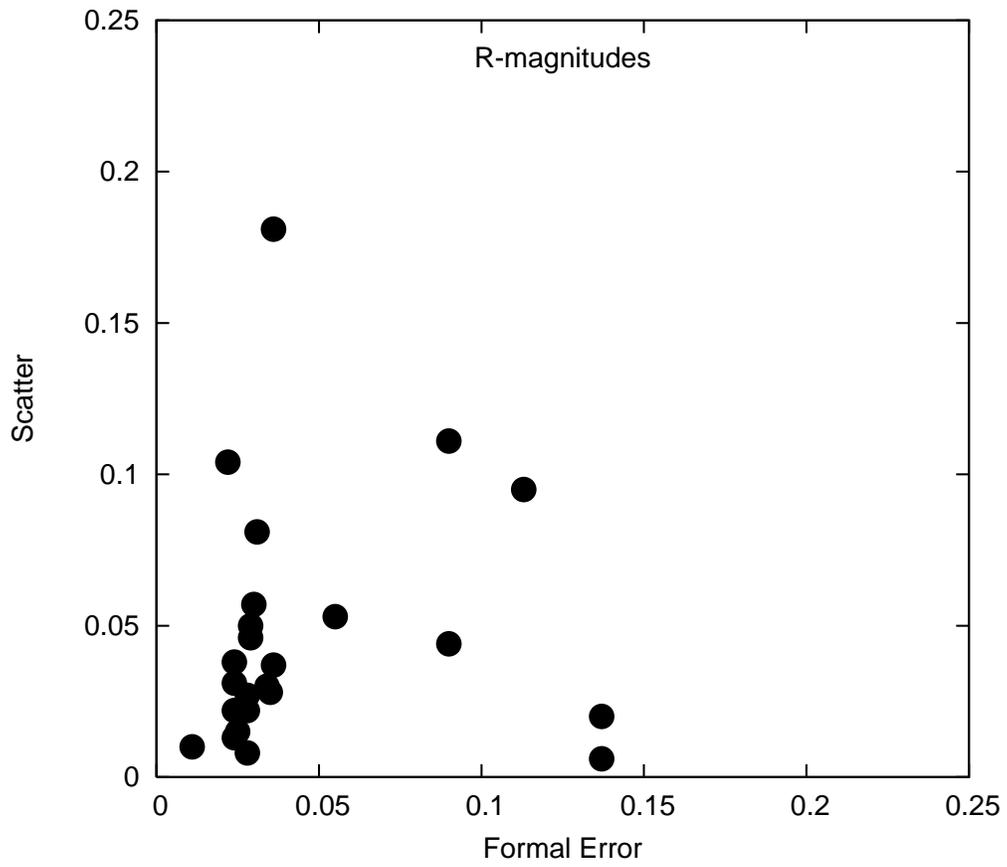}
\caption{Scatter in the {\it R} band data as a function of largest formal error.  
  \label{Rerrorvsscatter}}
\end{figure}

\begin{figure}
\plotone{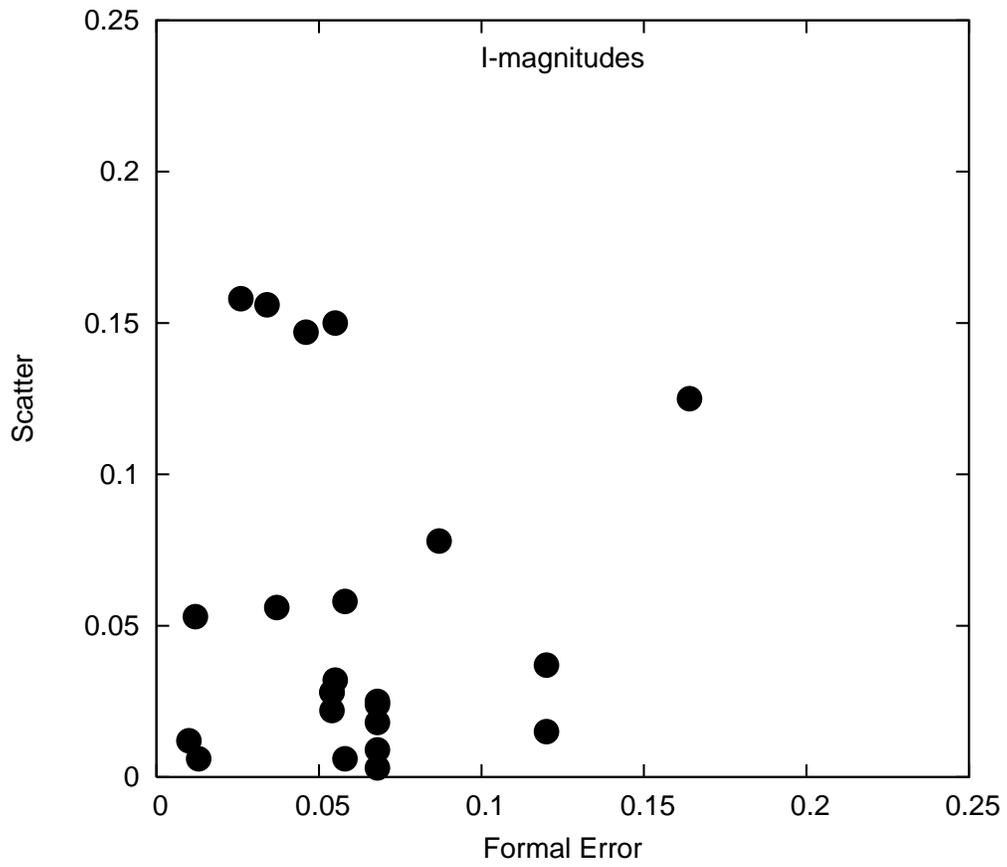}
\caption{Scatter in the {\it I} band data as a function of largest formal error.  
  \label{Ierrorvsscatter}}
\end{figure}

Figures~\ref{Berrorvsscatter} - \ref{Ierrorvsscatter} display our 
variability investigation results. For each filter ({\it B}, {\it V}, {\it R}, and 
{\it I}) the observed scatter (over different epochs) is plotted vs.~the largest, 
internal, formal error of all observed epochs.  
The strongest indication of variablity is seen for source 0552-640 
with {\it B} and {\it V} scatter of about 0.1 mag and formal error of 0.01 mag.
Several other sources show a ratio of scatter to formal error of
about 3, but more observations are required to characterize the 
photometric data for variability or draw any conclusions.

The largest observed scatter is about 0.2 mag over the few years
of our observing program which is in stark contrast to some of
the differences seen between the Veron-Cetty listed magnitudes
and our observations.

The errors in the photometric calibration of the data of a typical
acceptable night are on the order of 20 - 50 mmag for all
filters.  This holds for the least-squares fit error as well as
for the standard error of the derived photometric zero-point constant.
Thus the observed discrepancies between our observations and the
Veron-Cetty catalog, as well as the indication for variability
seem to be of physical nature.

\section{Conclusions/Summary}\label{summary}
A sample of extragalactic, compact sources, mainly QSOs have been 
observed to characterize the photometric stability of these sources to 
better than 5\%. Of the 134 sources which have {\it V}
magnitudes in the Veron \& Veron-Cetty catalog a difference of over
1.0 mag is found for the observed$-$catalog magnitudes for about 36\%
of the common sources, and 10 sources show over 3 mag
difference.
As expected, the largest problem in this context is the intrinsic 
photometric variability of these sources which will require multiple 
observations at different epochs to downselect  ``stable" candidates 
and furthermore will likely require additional observations
close to the epoch of a future mission.

Although this program is driven by SIM preparatory science goals, 
our observations are of general interest providing accurate 
magnitudes and colors for a large sample of QSO targets at current epochs.
An optical quasar monitoring program, as e.g.~proposed by J.Schramm 
(about 1980) is desirable.  The current paper mainly forms the 
baseline providing mean {\it B}, {\it V}, {\it R}, and {\it I} measures at one or few epochs. At least 
for the brighter subset of our candidates such a program could be undertaken 
in a collaboration with adequately equipped amateur astronomers. We 
plan to continue our photometric quasar monitoring program in order to 
obtain optical variability information on these targets.

\acknowledgments

\section{Acknowledgements}

We thank Hugh Harris of the USNO Flagstaff Station for many helpful suggestions. 
This research has made use of NASA's Astrophysics Data System 
Bibliographic Services. This research has made use of the NASA/IPAC 
Extragalactic Database (NED) which is operated by the Jet Propulsion 
Laboratory, California Institute of Technology, under contract with 
the National Aeronautics and Space Administration.

\clearpage

\clearpage

\clearpage

\clearpage

\clearpage

\clearpage

\clearpage

\clearpage

\clearpage

\clearpage

\clearpage

\clearpage

\clearpage

\clearpage

\input{tab1.tex}\label{observesummary}

\clearpage

\input{tab2.tex}\label{finalresults}

\clearpage

\input{tab3.tex}\label{magvariability}

\clearpage

\input{tab4.tex}\label{formalerrors}

\end{document}

%% file: tab1.tex
\begin{deluxetable}{crrr}
\tablecolumns{4}
\tablewidth{0pc}
\tablecaption{Summary of Observations}
\tablehead{
\colhead{Telescope} & \colhead{Epoch} & \colhead{Label} & \colhead{Nights}}
\startdata
NOFS & 2005.28 &  n51 & 4 \\
NOFS & 2005.66 &  n52 & 3 \\
NOFS & 2006.15 &  n53 & 3 \\
NOFS & 2007.53 &  n54 & 3 \\
CTIO & 2005.86 &  c02 & 4 \\
CTIO & 2006.29 &  c03 & 4 \\
CTIO & 2006.92 &  c05 & 12 \\
CTIO & 2007.15 &  c06 & 5 \\
\enddata
\label{observesummary}
\end{deluxetable}

%% file: tab2.tex
\begin{deluxetable}{cccccccccccccccccc}
\tablecolumns{18}
\tabletypesize{\tiny}
\tablecaption{Magnitudes of SIM Candidate Quasars}
\tablewidth{0pc}
\tablehead{
\colhead{Source} & \colhead{Mean B} & \colhead{Mean V} &
\colhead{Mean R} & \colhead{Mean I} & \colhead{Stdd B} &
\colhead{Stdd V} & \colhead{Stdd R} & \colhead{Stdd I} &
\colhead{B-V} & \colhead{V-R} & \colhead{R-I} &
\colhead{V-I} & \colhead{nB} & \colhead{nV} &  
\colhead{nR} & \colhead{nI} & \colhead{Epoch}}
\startdata
0000$-$128 & 15.646 & 15.516 & 15.312 & 14.645 & 0.007 & 0.007 & 0.003 & 0.008 & 0.130 & 0.205 & 0.667 & 0.871 & 2 & 3 & 3 & 3 & c05 \\
0001$+$113 & 16.354 & 16.155 & 15.894 & 15.244 & 0.028 & 0.014 & 0.015 & 0.015 & 0.199 & 0.261 & 0.650 & 0.911 & 1 & 1 & 1 & 1 & n52 \\
0001$+$113 & 16.443 & 16.396 & 16.066 & 15.346 & 0.001 & 0.027 & 0.003 & 0.029 & 0.047 & 0.330 & 0.720 & 1.050 & 2 & 1 & 2 & 1 & c05 \\
0006$+$437 & 14.851 & 14.621 & 14.402 & 13.850 & 0.029 & 0.020 & 0.021 & 0.019 & 0.230 & 0.219 & 0.552 & 0.771 & 1 & 1 & 1 & 1 & n52 \\
0010$-$063 & 17.438 & 16.555 & 15.973 & 15.367 & 0.029 & 0.019 & 0.004 & 0.001 & 0.882 & 0.582 & 0.606 & 1.188 & 4 & 2 & 4 & 2 & c05 \\
0010$-$063 & 17.774 & 16.275 & 15.683 & 14.993 & 0.029 & 0.014 & 0.014 & 0.015 & 1.499 & 0.592 & 0.690 & 1.282 & 1 & 1 & 1 & 1 & n52 \\
0017$+$816 & 17.605 & 16.524 & \ldots & 15.688 & 0.024 & 0.017 & \ldots & 0.017 & 1.081 & 9.900 & 9.900 & 0.836 & 1 & 1 & 0 & 1 & n52 \\
0028$+$311 & 15.467 & 15.218 & 14.971 & 14.634 & 0.029 & 0.020 & 0.021 & 0.019 & 0.249 & 0.247 & 0.337 & 0.584 & 1 & 1 & 1 & 1 & n52 \\
\enddata
\label{finalresults}
\end{deluxetable}
 

%% file: tab3.tex
\begin{deluxetable}{rcrrrrrrrrrrrrrrr}
\tablecolumns{17}
\tabletypesize{\scriptsize}
\tablewidth{0pc}
\tablecaption{Magnitude Variability}
\tablehead{
\colhead{} & \colhead{} & \multicolumn{3}{c}{B-mag} & \colhead{} & 
\multicolumn{3}{c}{V-mag} & \colhead{} & \multicolumn{3}{c}{R-mag} 
& \colhead{} & \multicolumn{3}{c}{I-mag} \\
\cline{3-5} \cline{7-9} \cline{11-13} \cline{15-17} \\
\colhead{Source} & \colhead{Runs} & \colhead{Mean} & \colhead{Error}\tablenotemark{a} & \colhead{Scatter}  
& \colhead{} & \colhead{Mean} & \colhead{Error} & \colhead{Scatter}
& \colhead{} & \colhead{Mean} & \colhead{Error} & \colhead{Scatter}
& \colhead{} & \colhead{Mean} & \colhead{Error} & \colhead{Scatter}}
\startdata
0504$-$297 & 3 & 16.234 & 0.032 & 0.032 & & 16.220 & 0.024 & 0.037 & & 16.155 & 0.024 & 0.031 & & 15.880 & 0.054 & 0.028 \\
0517$-$442 & 3 & 15.475 & 0.023 & 0.038 & & 15.219 & 0.027 & 0.021 & & 14.868 & 0.028 & 0.027 & & 14.353 & 0.068 & 0.009 \\
0534$-$603 & 3 & 18.108 & 0.153 & 0.027 & & 17.125 & 0.151 & 0.024 & & 16.485 & 0.137 & 0.006 & & 15.904 & 0.120 & 0.015 \\
0536$-$517 & 3 & 16.748 & 0.024 & 0.042 & & 16.210 & 0.027 & 0.039 & & 15.781 & 0.028 & 0.022 & & 15.189 & 0.068 & 0.025 \\
0552$-$532 & 3 & 15.938 & 0.023 & 0.023 & & 15.737 & 0.027 & 0.006 & & 15.361 & 0.028 & 0.008 & & 14.976 & 0.068 & 0.003 \\
0552$-$640 & 3 & 15.460 & 0.152 & 0.107 & & 15.245 & 0.151 & 0.099 & & 14.910 & 0.137 & 0.020 & & 14.539 & 0.120 & 0.037 \\
0556$-$027 & 3 & \nodata & \nodata & \nodata & & 20.386 & 0.025 & 0.152 & & 19.120 & 0.113 & 0.095 & & 18.257 & 0.046 & 0.147 \\
0559$-$504 & 3 & 15.105 & 0.024 & 0.017 & & 14.963 & 0.024 & 0.013 & & 14.816 & 0.024 & 0.013 & & 14.438 & 0.054 & 0.028 \\
\enddata
\tablenotetext{a}{The Largest Formal Error is shown}
\label{magvariability}
\end{deluxetable}
 

%% file: tab4.tex
\begin{deluxetable}{ccrrr}
\tablecolumns{5}
\tablewidth{0pc}
\tablecaption{Formal Error Statistics}
\tablehead{
\colhead{Filter} & \colhead{Number of}  & \colhead{Error} & \colhead{Median}
& \colhead{Largest} \\
\colhead{} & \colhead{objects}  & \colhead{$\le$ 5\%} & \colhead{error}
& \colhead{error}}
\startdata
 \hline
    B     &  130      & 107      &  0.021    & 0.44 \\
    V     &  128      & 102      &  0.020    & 0.54 \\
    R     &  142      & 112      &  0.018    & 0.56 \\
    I     &  131      & 100      &  0.025    & 0.60 \\
 \hline
\enddata
\label{formalerrors}
\end{deluxetable}